\documentclass[prd,onecolumn,groupedaddress, nofootinbib, 10pt]{revtex4}  
\usepackage{graphicx}
\usepackage{amsmath}
\usepackage{amsfonts}
\usepackage{amssymb}
\usepackage{appendix}
\usepackage{mathtools}
\usepackage{comment}
\usepackage{bbold}
\usepackage{color}
\usepackage{slashed}
\usepackage{subfigure}
\usepackage{setspace}
\usepackage{enumitem}
\usepackage{longtable}
\usepackage{wasysym}
\usepackage[usenames,dvipsnames]{xcolor}
\usepackage{bm}
\usepackage[letterpaper, margin=.8in, top=0.85in, bottom=0.85in]{geometry}

\pdfoutput=1

\begin{document}

\singlespacing

\title{Gravitational Effects on Measurements of the Muon Dipole Moments}
\author{Andrew Kobach}
\affiliation{Physics Department, University of California, San Diego, La Jolla, CA 92093, USA}

\date{\today}

\begin{abstract}
If the technology for muon storage rings one day permits sensitivity to precession at the order of $10^{-8}$ Hz, the local gravitational field of Earth can be a dominant contribution to the precession of the muon, which, if ignored, can fake the signal for a nonzero muon electric dipole moment (EDM).  Specifically, the effects of Earth's gravity on the motion of a muon's spin is indistinguishable from it having a nonzero EDM of magnitude $d_\mu \sim 10^{-29}$ e cm in a storage ring with vertical magnetic field of $\sim$ 1 T, which is significantly larger than the expected upper limit in the Standard Model, $d_\mu \lesssim 10^{-36}$ e cm.  As a corollary, measurements of Earth's local gravitational field using stored muons would be a unique test to distinguish classical gravity from general relativity with a bonafide quantum mechanical entity, i.e., an elementary particle's spin.  
\end{abstract}

\maketitle

\section{Introduction}

While colliders are suitable for producing heavy particles, uncovering precise detail of the underlying theory is left to precision measurements performed by dedicated experiments. This is particularly true for the electromagnetic interaction.  For example, the interaction between a charged lepton $\ell$ and an external electromagnetic field is 
\begin{equation}
\label{emint}
\left\langle \ell(p+q) \left| J^\alpha_\text{EM} (q^2) \right| \ell(p) \right\rangle \propto   \overline{u}(p+q) \bigg[  \gamma^\alpha F_1(q^2) + \frac{i\sigma^{\alpha\beta} q_\beta}{2m_\ell} F_2(q^2) - \sigma^{\alpha\beta}  q_\beta \gamma^5 F_3(q^2)  \bigg] u(p) ,
\end{equation}
where $m_\ell$ is the mass of the lepton and $\sigma^{\alpha\beta} \equiv \frac{i}{2}[\gamma^\alpha,\gamma^\beta]$. When $q^2=0$, $F_1(0)\equiv 1$, and the other form factors have values $F_2(0)= a_\ell$, $F_3(0)= d_\ell/e $, where $a_\ell$ is the anomalous magnetic dipole moment, and $d_\ell$ is the electric dipole moment (EDM).\footnote{There can be additional terms in Eq.~(\ref{emint}) due to toroidal moments, but their values are, in general, gauge dependent for elementary fermions~\cite{Musolf:1990sa}, and will not be considered in this analysis.  }  The definition of $F_1(q^2=0)$ is a requirement that the theory behaves like classical electrodynamics at low energies, and the values of $F_2$ and $F_3$ are purely quantum mechanical in origin and can be calculated in perturbation theory. 
Colliders can measure the gross scattering probability between charged particles, but they are generally insensitive to these higher-order effects.  
Dedicated experiments to measure the dynamics of a charged lepton's spin in external electromagnetic fields, on the other hand, can be singular opportunities to discover deviations from the Standard Model (SM), since theoretical calculations of charged-lepton dipole moments include all physics in the SM and, in principle, beyond.

Measurements of $a_\mu$, $d_e$, and $d_\mu$ serve as the best opportunities to discover new physics. 
On the other hand, while the electron is the easiest particle to control, experimentally, the potential to discover new physics with $a_e$ is minimized, because the value of $a_e$ is dominated by QED, and not the SM at the weak scale, due to the lightness of the electron mass~\cite{Aoyama:2012wj}.
Taus can be produced in appreciable amounts at collier experiments, but because of its short lifetime, neither $a_\tau$ nor $d_\tau$  have yet been measured with meaningful precision~\cite{Fael:2013ij, Eidelman:2016aih, Fael:2016hnz}.  

The muon anomalous magnetic dipole moment $a_\mu$ was best measured to $\sim 0.5$ ppm by the $(g-2)_\mu$ experiment at BNL~\cite{Bennett:2006fi}, but resulted in about a $2.7\sigma$ - $3.6\sigma$ discrepancy with the value expected in the SM, depending on the details of the contributions of the strong interactions~\cite{Davier:2010nc, Hagiwara:2011af, Kurz:2014wya, Prades:2009tw, Nyffeler:2009tw, Colangelo:2014qya}.  Whether or not this is a sign of new physics may be settled as improved experimental~\cite{Grange:2015fou, Iinuma:2011zz} and theoretical techniques~\cite{Boyle:2011hu, Aubin:2013daa, Chakraborty:2014mwa, Golterman:2014ksa, deRafael:2014gxa, Blum:2015you, Blum:2015gfa, Chakraborty:2015ugp, Chakraborty:2016mwy} are employed.   

The SM predictions for the values of $d_e$ and $d_\mu$ are minuscule, due to small magnitude of CP-invariance violation in the SM and the small values of neutrino masses.  The CP-invariance violation from mixing in the quark sector is the dominant contribution to charged-lepton EDMs,\footnote{Even if other sources of CP-invariance violation are present due from the mixing in the leptonic sector, their contribution to $d_\mu$ is very small~\cite{Ng:1995cs, deGouvea:2005jj, Pospelov:1991zt}.} where a nonzero muon EDM first occurs at the four loops, resulting in the upper limits $d_e \lesssim 10^{-38}$ e cm and $d_\mu \lesssim 10^{-36}$ e cm~\cite{Ng:1995cs, deGouvea:2005jj, Pospelov:1991zt}.  New physics, then, can easily play a dominant role over the SM in the contribution to the charged-lepton EDMs.  The current limits on the electron and muon EDMs are $d_e < 1.6\times10^{-27}  $ e cm (90\% CL) and  $d_\mu < 1.05\times 10^{-19}$ e cm (95\% CL)~\cite{Regan:2002ta, Bailey:1978mn}.  While this strong limit on $d_e$ can imply a stronger limit on the value of $d_\mu$ than what is currently experimentally verified, the naive linear scaling in the SM $d_e/d_\mu \sim m_e/m_\mu$ can be violated in the presence of new physics~\cite{Babu:2000cz, Feng:2001sq}.   New experimental techniques with stored muons~\cite{Farley:2003wt} offers opportunities for experiments to improve the limit on $d_\mu$ by almost 5 orders of magnitude, i.e., down to $d_\mu \lesssim 10^{-24}$ e cm, by choosing the electromagnetic fields such that the precession due to $a_\mu$ is minimized~\cite{Semertzidis:1999kv, Adelmann:2006ab}.  While further technological advances would be required to strengthen this limit further, such improvements would be invaluable to ruling out or discovering new physics that violates CP-invariance.  

A measurement of a nonzero muon EDM is often thought to be clear signal of new physics.  However, this might not always be the true.  There is a unique relativistic effect for particles in a cyclotron:~the local gravitational field will cause the muon's spin to precess in a way that mimics the dynamics due to a bonafide dipole moment.  
To date, no experiment has yet verified that the quantum mechanical spin of elementary particles obey general relativity.  It remains an experimentally open question whether non-composite massive quantum mechanical entities obey classical gravity or general relativity.  
In the following section, the expectation for the precession of a stored muon due to classical gravity and general relativity will be compared.  
For a typical cyclotron with a vertical magnetic field of $\sim$ 1 T, Earth's gravity will induce the same dynamics as a nonzero muon EDM of magnitude $d_\mu \sim 10^{-29}$ e cm, which is about 5 orders of magnitude greater than the upper limit on the SM expectation.

\section{Gravitational Effects on a Stored Muon}
\label{calc}


The Hamiltonian for a muon with charge $q$ an external electromagnetic field contains the following dipole terms:
\begin{equation}
\mathcal{H}_\text{dipole} = - \bm{s} \cdot \left( \frac{qg_\mu}{2m_\mu } \bm{B}' +  \frac{q\eta }{2m_\mu c} \bm{E}' \right),
\end{equation}
where $(g_\mu-2)/2 \equiv a_\mu$, and $\bm{d}_\mu \equiv \bm{s}(\eta q /mc)$.  Here, the primes indicate that these are the fields perceived by the particle in its rest frame. The equation of motion in the rest frame of the particle is 
\begin{equation}
\left( \frac{d\bm{s}}{d\tau} \right)_\text{rest} = \bm{s} \times \left( \frac{qg_\mu}{2m_\mu } \bm{B}' +  \frac{q\eta }{2m_\mu c} \bm{E}' \right) .
\end{equation}
These electromagnetic fields are however not the fields in the (unprimed) lab frame:
\begin{eqnarray}
\bm{B}' &=& \gamma \left( \bm{B} - \frac{\bm{\beta}\times \bm{E}}{c} \right) - \left(\frac{\gamma^2}{\gamma+1}\right) \bm{\beta} (\bm{\beta} \cdot \bm{B})  \\
\bm{E}' &=& \gamma \left( \bm{E} + c\bm{\beta}\times \bm{B} \right) - \left(\frac{\gamma^2}{\gamma+1}\right) \bm{\beta} (\bm{\beta} \cdot \bm{E}) 
\end{eqnarray}

The equation of motion a particle's spin in the lab frame  differs depending on the theory of gravity used.  Classically, for example, gravity is just a force like any other, where the muon's spin equation of motion is calculated by considering only the noncommutativity of Lorentz transformations, which is an effect described by the well-known Thomas precession of angular momentum:
\begin{equation}
\label{dsdt}
\left( \frac{d\bm{s}}{dt} \right)_\text{lab} = \bm{\omega}_s \times \bm{s} =  \left( \frac{d\bm{s}}{d\tau} \right)_\text{rest}  - \left[\left(\frac{\gamma^2}{\gamma+1}\right) \frac{ \bm{\beta} \times \bm{a}}{c} \right] \times \bm{s} , 
\end{equation}
where $t$ is the time in the lab frame.
Here, $\bm{a}$ is the classical acceleration of the particle to the local electromagnetic and gravitational fields. To calculate the acceleration, one can note that the total energy associated with the particle's motion is $\mathcal{E} = \gamma m c^2$, and the force in the lab frame is therefore
\begin{eqnarray}
\frac{d\bm{p}}{dt} = \frac{d(\gamma m_\mu  \bm{v})}{dt} = \frac{d}{dt} \left( \frac{\mathcal{E} \bm{v}}{c^2} \right)  = \frac{\bm{v}}{c^2} \bigg[ q(\bm{v} \cdot \bm{E}) + m_\mu (\bm{v} \cdot \bm{g})\bigg] + \frac{\mathcal{E}}{c^2} \frac{d\bm{v}}{dt},
\end{eqnarray}
and also that the force on the particle in the lab frame is due to the gravitational and electromagnetic fields:
\begin{equation}
\label{classical}
\frac{d\bm{p}}{dt} = m_\mu \bm{g} + q\left( \bm{E} + \bm{v} \times \bm{B} \right) .
\end{equation}
Noting that $\bm{a} \equiv d\bm{v}/dt$ and $\bm{\beta} \equiv \bm{v}/c$, the acceleration is: 
\begin{equation}
\label{accel}
\bm{a} = \frac{q}{m_\mu \gamma} \big( \bm{E} + c \bm{\beta} \times \bm{B} - \bm{\beta} (\bm{\beta}\cdot \bm{E}) \big) + \frac{1}{\gamma}\big( \bm{g} - \bm{\beta} (\bm{\beta} \cdot \bm{g}) \big),
\end{equation}
where $\bm{g}$ is the gravitational acceleration at the surface of Earth.

On the other hand, the equation of motion for the muon's spin in general relativity takes into account the effects of spacetime's curvature.  In this case, such an effect was calculated in Refs.~\cite{Khriplovich:1997ni, Silenko:2004ad} for a weak gravitational field:
\begin{equation}
\label{GR}
\left( \frac{d\bm{s}}{dt} \right)_\text{lab} = \bm{\omega}_s \times \bm{s} =  \left( \frac{d\bm{s}}{d\tau} \right)_\text{rest}  - \left[ \left(\frac{\gamma^2}{\gamma+1}\right) \frac{ \bm{\beta} \times \bm{a}}{c} - \left(\frac{2\gamma+1}{\gamma+1}\right) \frac{\bm{\beta} \times \bm{g}}{c} \right] \times \bm{s}, 
\end{equation}
Here, $\bm{a}$ is  the acceleration due to only the external electromagnetic fields, i.e., the expression for $\bm{a}$ in Eq.~(\ref{accel}) without the terms involving $\bm{g}$. 

Given either Eqs.~(\ref{classical}) or (\ref{GR}), the precession of the muon's spin is described by the angular velocity
\begin{eqnarray}
\label{omegas}
\bm{\omega}_s &=& - \frac{q}{m_\mu } \bigg[ \left( \frac{g_\mu}{2} - 1 + \frac{1}{\gamma} \right) \bm{B} + \left( \frac{\gamma}{\gamma+1} - \frac{g_\mu}{2}  \right) \frac{\bm{\beta}\times\bm{E}}{c} + \frac{\gamma}{\gamma+1}\left(1-\frac{g_\mu}{2} \right) \bm{\beta}(\bm{\beta}\cdot \bm{B}) \nonumber \\
&& \qquad + ~ \frac{\eta}{2c} \big( \bm{E} + c\bm{\beta}\times\bm{B} - \bm{\beta} (\bm{\beta}\cdot \bm{E}) \big)  \bigg] + \bm{\omega}_g  .
\end{eqnarray}
Eq.~(\ref{omegas}) is the well-known precession frequency for the muon's spin in constant external electromagnetic fields, plus an extra term due to the effects of Earth's gravity. Here, if the velocity of the muon is perpendicular to the direction toward the center of Earth:
\begin{equation}
\label{omegag}
|\bm{\omega}_g| = \left| \left( \frac{\gamma+\kappa(\gamma+1) }{\gamma+1} \right) \frac{\beta g}{c} \right|
\end{equation}
The classical-gravity prediction corresponds to when $\kappa = 0$, and the prediction using general relativity is associated with $\kappa = 1$ (note that there is no physical significance to the values  $0<\kappa<1$).  The precession in either scenario due to gravity does not depend on the particle's mass nor the external electromagnetic fields, rather only the muon's velocity.  This is because the effect is purely relativistic, due to the nontrivial dynamics of angular momentum in accelerating reference frames.  In the limit when $\gamma \rightarrow 1$, the result using general relativity for $|\bm{\omega}_g|$ is a factor of 3 larger than the one using just classical gravity, which is due to the effects of spatial curvature~\cite{Mashhoon:2013jaa}.

Most cyclotron rings are constructed in a plane, where $\bm{B}$ is vertical,\footnote{Cyclotrons are constructed such that the magnetic field along the ring is parallel to Earth's gravitational force, e.g., see Ref.~\cite{Grange:2015fou}. } $\bm{E}$ is radial, and $\bm{\beta}$ is azimuthal.  Therefore, since $\bm{g}$ is also vertical, the gravitational term in Eq.~(\ref{omegas}) can mimic the motion caused by a non-zero EDM.  When $\beta\approx 1$ and $q=e$ ($e$ being the fundamental electric charge), the magnitude of the EDM that contributes the same amount as gravity to the precession of the particle is
\begin{equation}
\label{result}
\eta \sim \frac{2m_\mu |\bm{g}|}{ec|\bm{B}|} \hspace{0.25in}\Longrightarrow \hspace{0.25in} d_\mu \sim (7\times10^{-30} \text{ e cm}) (\kappa+1)\left( \frac{\text{1 T}}{|\bm{B}|}\right).
\end{equation}
Note that the two EDM terms in Eq.~(\ref{omegas}) that depend on $\bm{E}$ are negligible compared to the one that depends on $\bm{B}$.  Again, here $\kappa=0$ (1) corresponds to classical (general relativistic) predictions for the precession effects due to gravity.  The magnitude of the precession rate in Eq.~(\ref{result}) is similar to that in Ref.~\cite{Silenko:2006er} when general relativity was used in the presence of magnetic focusing for stored deuterons.  An important point to emphasize is that such spin precession due to gravity for stored muons will occur in the contexts of both classical gravity and general relativity, their difference being an important factor of 2 in precession rate.

\section{Summary and Conclusions}

With the strong possibility to discover or rule out new physics beyond the SM, significant motivation exists to execute the most precise  measurements possible of the values of charged-lepton dipole moments.  Such precision, however, comes at the price of understanding in minute detail the expected behavior of the experiment on a whole.  One such effect is the role of gravity in these experiments.  In particular, gravity will produce a non-trivial effect in relativistic systems with rotating references frames.  This effect can manifest itself in muon storage rings, where the local gravitational field alone can cause the muon's spin to precess.  As calculated in Sec.~\ref{calc}, the effect of Earth's gravity on the precession of a muon (corresponding to a precession rate of $\sim 10^{-8}$ Hz) will fake the signal for a nonzero EDM of $d_\mu \sim 10^{-29}$ e cm in a storage ring with a vertical magnetic field of magnitude $|\bm{B}|\sim$ 1 T.  This gravitational effect fakes an muon EDM that is much larger than the SM upper limit $d_\mu \lesssim 10^{-36}$ e cm.

The storage ring technology needed to gain this level of precision would require significant technological advances beyond the current state-of-the-art experiments.  But if such ability is one day obtained, there are at least two interesting issues to be addressed.  First, if one wishes to gain experimental precision of the precession rate of a relativistic muon in an external electromagnetic field beyond $\sim10^{-8}$ Hz, then very precise information would be needed regarding the local gravitational effects at the storage ring.  Second, if no new physics is assumed to contribute to the value of $d_\mu$, then it would be a singular opportunity for gravity to be measured by a fundamental particle's spin, testing the predictions of special and general relativity with a bonafide quantum mechanical entity.   In particular, it would provide major insight to discover whether an elementary particle's spin obeys classical gravity or general relativity, which is yet an unverified experimental fact.

\begin{acknowledgments}
AK is grateful for useful comments and feedback from Andr\'{e} de Gouv\^{e}a, Ben Grinstein, and Heidi Schellman.  The work of AK is funded in part by DOE grant \#DE-SC0009919. 
\end{acknowledgments}

\bibliography{bib}{}

\end{document}